\documentclass[twocolumn,showpacs,aps,prd]{revtex4}

\usepackage{graphicx}
\usepackage{dcolumn}   
\usepackage{amsmath} 
\usepackage{epsfig}
\usepackage{hhline}
\RequirePackage{xspace}
\usepackage{bm}

\def\sbar{{\overline s}}
\def\figurebox#1#2#3{%
    \def\arg{#3}%
    \ifx\arg\empty
    {\hfill\vbox{\hsize#2\hrule\hbox to #2{\vrule\hfill\vbox to #1{\hsize#2\vfill}\vrule}\hrule}\hfill}%
    \else
    {\hfill\epsfbox{#3}\hfill}%
    \fi}

\newcommand{\bsigma}{\bm{\sigma}}
\newcommand{\rhat}{\hat{\bm r}}
\def\ket#1#2#3{|#1,\,#2,\,#3\rangle}

\newcommand{\non}{\nonumber\\}
\newcommand{\ncdot}{\!\cdot\!}
\begin{document}

\preprint{LBNL-xxxx}
\begin{flushleft}
LBNL-52572
hep-ph/
\end{flushleft}

\title{
{\large \bf Spin-Orbit and Tensor Forces in Heavy-quark Light-quark Mesons:
Implications of the New $D_s$ State at 2.32 GeV
%
}
}

\author{ Robert N. Cahn and J. David Jackson}
\affiliation{Lawrence Berkeley National Laboratory\\ 1 Cyclotron Rd.,\\
Berkeley, CA 94720)}
\date{\today}

\begin{abstract}
We consider the spectroscopy of heavy-quark light-quark mesons with a
simple model based on the non-relativistic reduction of vector and
scalar exchange between fermions.  Four forces are induced: the
spin-orbit forces on the light and heavy quark spins, the tensor
force, and a spin-spin force.  If the vector force is Coulombic, the
spin-spin force is a contact interaction, and the tensor force and
spin-orbit force on the heavy quark to order $1/m_1m_2$ are directly
proportional.  As a result, just two independent parameters
characterize these perturbations.  The measurement of the masses of
three p-wave states suffices to predict the mass of the fourth.  This
technique is applied to the $D_s$ system, where the newly discovered
state at 2.32 GeV provides the third measured level, and to the $D$
system. The mixing of the two $J^P=1^+$ p-wave states is reflected in
their widths and provides additional constraints.  The resulting picture is
at odds with previous expectations and raises new puzzles.

\end{abstract}

\maketitle

\newpage

\setcounter{footnote}{0}
\section{\boldmath Introduction}
Mesons composed of one light quark and one heavy quark are quite
analogous to the hydrogen atom \cite{dgg} and can be analyzed using
the traditional methods \cite{bethe-salpeter}.  The fine and hyperfine
structures have direct analogues in meson spectroscopy, but the
confinement of quarks requires that potential cannot be purely
Coulombic.  A convenient phenomenological approach is to postulate
that there are two separate static potentials, one arising as the zeroth
component of a vector potential and the other as a Lorentz scalar.
Asymptotic freedom suggests that the vector potential be Coulombic,
while confinement suggests that the scalar be a linear potential.
Such models have had reasonable success in describing the spectroscopy
of the $D$, $D_s$, $B$, and $B_s$ systems \cite{godfrey-isgur,
godfrey-kokoski, isgur-wise, eichten-hill-quigg, dipierro-eichten}.

The recently discovered state with a mass of 2.32 GeV decaying to $D_s\pi^0$
\cite{babar} appears to be a $c{\overline s}$ p-wave meson with $J^P=0^+$.  
Typical predictions for its mass were near 2.5 GeV, above the
threshold for the strong decay to $D K$.  Here we analyze the p-wave
$c {\overline s}$ and $c {\overline u}/{\overline d}$ mesons in a
simple model that is more general, though less predictive, than the
models described above.  It is more general in that we do not
specialize to a scalar potential that is linear. 

Our concern here is restricted to the p-wave states.  While our primary
interest is in the $D_s$ system, we consider as well the analogous 
non-strange $D$ mesons.  It is not only the mass spectrum that needs to be
addressed, but also the pattern of decay widths.  The early studies, which
predicted a much higher mass for the $J^P=0^+$ $c\sbar$, were successful in
explaining the narrow width of the observed $J^P=1^+$ states in both
the $D$ and $D_s$ systems.  We find that the combined constraints of the
mass spectrum as now known (though lacunae remain)  and the decay patterns make it difficult to find a consistent
picture of the p-wave charmed strange and non-strange mesons.

By analogy with the hydrogen atom, a convenient classification of
states is given in terms of ${\bm j}={\bm\ell}+{\bm s}_1$, where ${\bm
\ell}$ is the orbital angular momentum and ${\bm s}_1$ is the spin of
the light quark.  The light-quark angular momentum ${\bm j}$ is
conserved in the limit in which the heavy-quark mass, $m_2$, goes to
infinity.  For p-wave states, the values of $j$ are $3/2$ and $1/2$.
These levels are split by the ordinary spin-orbit force.  The
hyperfine structure is proportional to $1/m_1m_2$ and includes the
spin-orbit coupling of the heavy quark, a spin-spin interaction, and
the tensor force.  All these contribute to further reducing the
degeneracy so that ${\bm J}={\bm j}+{\bm s}_2$ is conserved, but not
${\bm j}$ alone.  The result is four distinct states with $J^P= 0^+,
1^+, 1^+,$ and $2^+$.  The two $J^P=1^+$ states are mixtures of
$j=3/2$ and $j=1/2$.  The $J=2$ and $J=0$ states are pure $S=1$
states, where ${\bm S}={\bm s}_1+{\bm s}_2$ is the total spin.  The
$J=1$ eigenstates are mixtures of spin-triplet and spin-singlet.

The p-wave $D$ mesons decay by pion emission.  The $2^+$ state decays
through d-wave emission to the ground state $0^-$, $D$ or to the
 $1^-$, $D^*$.  The $1^+$ states can decay, in principle, either by
d-wave or s-wave pion emission.  The limited phases space favors the
s-wave decay.  However, the decay of $j=3/2$ state to the $D^*$, which
has $j=1/2$, is forbidden to the extent that ${\bm j}$ is conserved.
Indeed, the $J^P=1^+$ state at 2.422 GeV is narrow.

For p-wave $D_s$ mesons, decay by pion emission is forbidden by
isospin invariance.  The $D_s$ states at 2.572 GeV and 2.536 GeV decay
to $DK$ and $D^*K$.  In a fashion analogous to the pattern in the $D$
system, the $J^P=1^+$ state at 2.536 GeV is also narrow. 

\section{\boldmath Spectrum for P States}
We base our analysis on 
 the quasi-static potential, including the
spin-dependent forces, which computed from the Lorentz-invariant fermion-antifermion
scattering amplitude using Feynman diagrams, replacing the vector or
scalar propagators by the Fourier transforms of the postulated potentials, $S(r)$ and $V(r)$. In this way, neglecting velocity-dependent, but spin-independent terms, we find
\begin{widetext}
\begin{eqnarray}
{\cal V}_{quasi-static}&=& V + S+\left({V'-S'\over r}\right){\bm \ell}\ncdot\left({\bsigma_1\over 4m_1^2}+{\bsigma_2\over 4m_2^2}\right)
+\left({V'\over r}\right){\bm \ell}\ncdot\left({\bsigma_1+\bsigma_2\over 2m_ 1m_2}\right)\nonumber\\[0.1in] 
&&\qquad\qquad\qquad\qquad+{1\over 12 m_1m_2}\left({V'\over r}-V''\right)S_{12}
+{1\over 6m_1m_2}\nabla^2 V \,\bsigma_1\ncdot\bsigma_2
\end{eqnarray}
\end{widetext}
where the tensor force operator is
\begin{equation}
S_{12}=3\,\bsigma_1\ncdot\rhat\,\bsigma_2\ncdot\rhat -\bsigma_1\ncdot\bsigma_2.
\end{equation}

We imagine solving the corresponding differential equation with the
potential $V+S$, then computing the fine structure and hyperfine structure
perturbatively.  We work consistently only to order $1/m_2$.  In general,
there are four independent matrix elements to consider, $\langle (V'-S')/r\rangle$,  $\langle V'/r\rangle$,    $\langle -V''+ V'/r\rangle$, and
$\langle\nabla^2V\rangle$.  However, if $V$ is Coulombic, $-V''+V'/r=3V'/r$
and  $\langle\nabla^2V\rangle$ vanishes except for s waves, where it gives 
a contact interaction.  It follows that for Coulombic $V$ and for $\ell>0$,
there are only two independent matrix elements.  Three measured p-wave masses
will give two splittings, which will determine the matrix elements and allow
us to predict the mass of the fourth p-wave state.

The mass operator for the splittings of the four states of any multiplet of orbital angular
momentum different from zero can be written (to order $1/m_2$)\cite{rosner}
\begin{equation}
M=\lambda {\bm \ell}\cdot {\bm s}_1+4\tau{\bm \ell}\cdot{\bm s}_2 + \tau S_{12}
\label{eq:ansatz}\end{equation}
with the notation
\begin{eqnarray}
\lambda&=&{1\over 2m_1^2}\left[{V'\over r}\left(1+{2m_1\over m_2}\right)-{S'\over r}\right]\non
\tau&=&{1\over 4m_1m_2}{V'\over r}.\label{eq:lambdatau}
\end{eqnarray}
In practice, we shall apply this operator to p-wave states for the $D$ or $D_s$ system.
Henceforth, we will use $\lambda$ and $\tau$ to indicate the expectation values of
the quantities in Eq.~(\ref{eq:lambdatau}).  The values of $\lambda$ and $\tau$ will
be different for the $D$ and $D_s$ systems.
  For the
assumed attractive, Coulombic $V$, the tensor-force energy $\tau$ is manifestly positive.  However, the spin-orbit energy $\lambda$ can be either positive or
negative, depending on the relation between the potentials $V$ and $S$.

Because there is no single basis that diagonalizes all the interactions,
we need to fix one basis, then calculate the mixing between the two
$J=1$ states.  We choose the basis in which $j^2$ is diagonal. The eigenstates  $\ket Jjm$
of $J^2$, $j^2$ and $J_z$ can be written in terms of eigenstates of
$J^2$, $S^2$, and $J_z$, or in terms of eigenstates of $J^2$, $(\bm j')^2=(\bm\ell +\bm s_2)^2$ and $J_z$.  The details are given in the Appendix.

\begin{figure}[b]\begin{center}
\includegraphics[width=3in,angle=90]{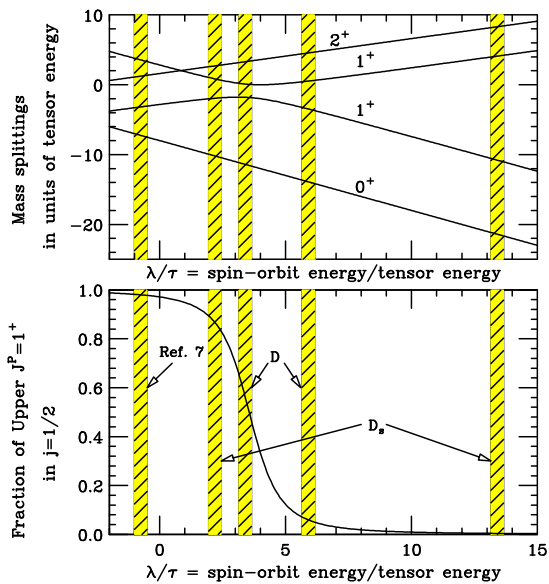}
\caption{Upper: the energy-levels for the four p-wave states as a function
of the ratio of the spin-orbit to tensor energies, in units of the tensor
energy.  Lower: the percentage of the more massive $J^P=1^+$ state that comes
from the $j=1/2$ state. The $D$ masses are from \cite{rpp, belle} and the 
$D_s$ masses from \cite{rpp,babar}.\label{fig:one}}
\end{center}\end{figure}

Up to an additive constant common to the four p-wave states, the masses of the
$J=2$ and $J=0$ states are
\begin{eqnarray}
M_2= {\lambda\over 2}+{8\over 5}\tau;\qquad
M_0= -\lambda -8\tau
\end{eqnarray}
while the masses of the two $J=1$ state are obtained by diagonalizing
the matrix in the $\ket Jjm$ basis
\begin{equation}
\left(\begin{array}{cc}
{\lambda\over 2}-\frac 83\tau & -\frac{2\sqrt 2}3\tau\\[0.2in]
-\frac{2\sqrt 2}3\tau        & -\lambda +\frac 83 \tau
\end{array}\right)
\end{equation}
The two eigenmasses for $J=1$ are then
\begin{eqnarray}
M_{1\pm}&=&-\frac{\lambda}4\pm\sqrt{\frac{\lambda^2}{16}+\frac 12(\lambda-4\tau)^2}\end{eqnarray}
The eigenmasses are shown as functions of $\lambda/\tau$ in Fig.~\ref{fig:one}.
Also shown is the $j=1/2$ fraction of the higher mass $J^P=1^+$ state.
The vertical bands correspond to theoretical models and to data, as explained in the next section.

If we define the mass splittings among three of the four states,

\begin{eqnarray}
D_2&=&M_2-M_0\non
D_1&=&M_{1}-M_0
\end{eqnarray}
we find 
\begin{eqnarray}
\tau&=& \frac{10}{87}D_2-\frac 2{29}D_1\nonumber\\
&&\qquad\pm\sqrt{\left(\frac{10}{87}D_2-\frac 2{29}D_1\right)^2
+\frac 5{232}(D_1^2-D_1D_2)}\nonumber\\
\lambda&=&\frac 23D_2-\frac{32}5\tau
\end{eqnarray}

  A very
strong scalar potential $S$ leads to ``inversion,'' namely the $j=1/2$ states lying above the $j=3/2$ states.

\section{\boldmath Application to the $D$ and $D_s$ Systems}

Using the formulae of the preceding Section, we can use the measured
masses of three p-wave states to predict the fourth.  Alternatively,
we can take three masses that are predicted theoretically and confirm
that we find the fourth predicted mass.  Applying this to the $D$ and
$D_s$ systems as given in Ref.~\cite{dipierro-eichten} we indeed find
congruence and  determine the values of $\lambda$ and $\tau$ shown
in Table~\ref{table:spectrum}.  The leftmost vertical bar in
Fig.~{\ref{fig:one}} indicates the range found in
Ref.~\cite{dipierro-eichten}.  The negative value of $\lambda$ is
indicative of inversion.  

The result of inversion is that the
higher-mass $J^P=1^+$ state actually lies above the $2^+$ state.  This
prediction is contradicted by the reports from Belle \cite{belle} of
$J^P=1^+$ state at 2.400 GeV.  Using the well established $D$ states
at 2.459 GeV and 2.422 GeV, and the reported $J^P=0^+$ at 2.290 GeV in
our Eq.~(\ref{eq:ansatz}) leads to two solutions, labeled A and B in
Table~\ref{table:spectrum}.  Both give masses near the observed state
at 2.400 GeV.  From the lower graph in Fig.~\ref{fig:one} we see that
the solution with the higher value of $\lambda/\tau$ results in the
higher-mass $J^P=1^+$ state (2.422 GeV) being nearly entirely $j=3/2$, while
the lower value (Solution A), would give it nearly equal contributions from
$j=1/2$ and $j=3/2$.

\begin{table}
\caption{Masses of the various p-wave states in the $D$ and $D_s$ systems and the
spin-orbit and tensor energies.
The experimental masses for the $D$ states at 2.400 GeV and 2.290 GeV are from
Ref.~\cite{belle}.  The mass of the $D_s$ state at 2.317 GeV is from 
Ref.~\cite{babar}.  The values of $\lambda$ and $\tau$ for Ref.~\cite{dipierro-eichten} were obtained by fitting their mass spectrum with the ansatz of
Eq.~\ref{eq:ansatz}. The square brackets indicate values that were used as inputs to the fits that determined the remaining mass and the values of $\lambda$ and $\tau$.  At the time of Ref.~\cite{dipierro-eichten} only the masses of the $2^+$ states and the 
$D_1(2420)$ and $D_{s1}(2535)$ were known. \label{table:spectrum}}  
\begin{tabular}{lrrrr}\hline
&\hskip 0.10in   Exp.\hskip 0.10in  &\multicolumn{3}{c}{Theory}\\ 
& Ref.~\cite{rpp,belle,babar}&Sol. A&Sol. B&Ref.~\cite{dipierro-eichten}\\ 
\hline\hline
\multicolumn{5}{l}{$D$ mesons}\\ \hline
$M(2^+)  $(GeV)&2.459&[2.459]&[2.459]&2.460\\
$M(1^+)  $(GeV)&2.400&2.400  &2.385&2.490\\
$M(1^+)  $(GeV)&2.422&[2.422]&[2.422]&2.417\\
$M(0^+)  $(GeV)&2.290&[2.290]& [2.290]&2.377\\  
$\lambda$ (MeV) &&39&54&$-11$\\
$\tau $ (MeV)   &&11&9&11\\ \hline
\multicolumn{5}{l}{$D_s$ mesons}\\ \hline
$M(2^+)  $(GeV)&2.572&[2.572]&[2.572]&2.581\\
$M(1^+)  $(GeV)&    &2.480&2.408&2.605\\
$M(1^+)  $(GeV)&2.536&[2.536]  &[2.536]&2.535\\
$M(0^+)  $(GeV)&2.317&[2.317]& [2.317]&2.487\\  
$\lambda$ (MeV) &&43&115&$-7$\\
$\tau $ (MeV)   &&20&9&11\\ \hline
\end{tabular}

\end{table}

We can differentiate between the two solutions, both for $D$ and $D_s$ by
considering the widths.  The experimental widths and theoretical estimates
are shown in Table~\ref{table:widths}.  The theoretical estimates are 
obtained from \cite{dipierro-eichten} after making phase-space corrections.
The widths for s-wave and d-wave decay are shown separately for the $J^P=1^+$
states.  The proper combination depends on the mixing of the $j=3/2$ and
$j=1/2$ states.
\begin{table}\begin{center}
\caption{\label{table:widths}Decay widths of p-wave $D$ and $D_s$ states in MeV.
The theoretical values are derived from \cite{dipierro-eichten} using
phase-space corrections to adjust for the masses known now. The widths shown for
the $D_2^*(2460)$ states are obtained from the total width of $23\pm5$ $0^-$ 
and the measured \cite{rpp} ratio  $\Gamma(D^+\pi^-)/\Gamma(D^{*+}\pi^-)=2.3\pm 0.6$.  For the $D_{sJ}(2573)$ we have assigned the entire width to the
decay to $D(1865)K$ since the decay to $D^*(2007)K$ has not been seen. }
\begin{tabular}{lrrr}\hline
&Exp. & \multicolumn{2}{c}{Theory:}\\ 
&\cite{rpp,belle}&s-wave&d-wave\\ \hline\hline
\multicolumn{4}{l}{$D$ mesons}\\ \hline
$D_2^*(2460)\to D(1865)\pi$&$16\pm4$&&16\\
$D_2^*(2460)\to D^*(2007)\pi$&$7\pm3$&&9\\
$D_1(2422)\to D^*(2007)\pi$&$18.9{+4.6\atop -3.5}$&90&10\\
$D_1(2400)\to D^*(2007)\pi $&$380\pm100\pm100$&100&\\
$D_0^*(2290)$&$305\pm30\pm25$&100&\\ \hline
\multicolumn{4}{l}{$D_s$ mesons}\\ \hline
$D_2^*(2573)\to D(1865) K$&$15{+5\atop -4}$&&9\\
$D_2^*(2573)\to D^*(2007) K$&$-$&&1.4\\
$D_1(2535)\to D^*(2007)K$&$<2.3$&100&0.3\\
 \hline
\end{tabular}
\end{center}\end{table}

Referring to Fig.~\ref{fig:one} and Table~\ref{table:spectrum}, we see
that the model of \cite{dipierro-eichten}, which has $\lambda<0$, has
the $J=1$ state that is nearly pure $j=1/2$ above the $J=2$ state.
The lower $J=1$ state is identified with the $D_1(2420)$.  Because it
is nearly entirely $j=3/2$ its decay by pion emission must be d-wave.
This causes a suppression that conforms to the measured width.
However, the mass spectrum predicted by Ref.~\cite{dipierro-eichten}
is not in good agreement with the data.  Of our two solutions for the
$D$ data, only one has a large value of $\lambda/\tau$ as needed to
make the lower mass $J=1$ state broad, as required by the data.

Turning to the $D_s$ system, we find a similar situation.  While
Ref.~\cite{dipierro-eichten} would have $\lambda<0$, the known masses,
which now include the $0^+$ state at 2.32 GeV, require $\lambda>0$.
Indeed, to suppress the width of the $D_s(2535)$, a large $\lambda$ is
needed. This is the case for Solution A, where the mass predicted for
the fourth p-wave state is 2408 MeV.  One of us (JDJ) notes that
in Ref.~\cite{babar},
there is an apparent signal in the invariant mass of $D_s\pi^0\gamma$
2.46 GeV.  This would be consistent with a $J^P=1^+$ state decaying
through $D_{sJ}^*(2317)\gamma$ or through $D^*_s(2112)\pi^0$.  The
mass fits better with Solution A of Table~\ref{table:spectrum}, but
the narrow width of the $D_{s1}(2535)$ favors Solution B.

Both the $D$ and $D_s$ systems show deviations from the pattern
anticipated by potential models.  While ``inversion'',
i.e. $\lambda<0$, has been a favored prediction, it is not in
agreement with the data.  This suggests that the ansatz taken for the
potentials $V$ and $S$ may not be as simple as assumed.  We have
allowed $S$ to be any potential, while requiring that $V$ be
Coulombic; Ref.~\cite{dipierro-eichten} required that $S$ be linear.
Relaxing that restriction might lead to improved agreement with the
data for this highly predictive approach, which actually uses the
solutions to the Dirac equation.  On the other hand, even with the
flexibility allowed in our approach, the resulting picture is not
entirely attractive.  We expect the confining potential to be quite
important since the p-wave states are not concentrated at the origin.
This would lead to a suppression of $\lambda$ through the contribution of
the $-S'/r$ term.  This is not borne out by the values of $\lambda$ we
deduce.

The discovery of the $D_{sJ}^*(2317)$ has provided an important clue to
heavy-quark light-quark spectroscopy by nailing down a p-wave state with
$j=1/2$.  Puzzles remain.  The anticipated discovery of the accompanying
$j=1/2$ state with $J=1$ should add important new information, but it
is not likely to resolve all the questions we have described.

 \begin{widetext}
\section{\boldmath Appendix}
The relations betweeen the three bases that diagonalize $j^2$, $j'^2$, and $S^2$ are
\begin{equation}
\begin{array}{l@{\ =\ }l}
\ket {J=\ell+1}{j=\ell+1/2}m& \ket {J=\ell+1}{S=1}m\nonumber\\[0.1in]
\ket {J=\ell}{j=\ell-1/2}m & \sqrt{{J+1\over 2J+1}}\,\ket {J=\ell}{S=1}m - \sqrt{{J\over 2J+1}}\,\ket {J=\ell}{S=0}m\nonumber\\[0.1in]
\ket {J=\ell}{j=\ell+1/2}m & \sqrt{{J\over 2J+1}}\,\ket {J=\ell}{S=1}m  + \sqrt{{J+1\over 2J+1}}\,\ket {J=\ell}{S=0}m\nonumber\\[0.1in]
\ket {J=\ell-1}{j=\ell-1/2}m & \ket {J=\ell-1}{S=1}m\\[0.15in]
\ket {J=\ell+1}{j'=\ell+1/2}m & \ket {J=\ell+1}{S=1}m\nonumber\\[0.1in]
\ket {J=\ell}{j'=\ell-1/2}m & \sqrt{{J+1\over 2J+1}}\,\ket {J=\ell}{S=1}m + \sqrt{{J\over 2J+1}}\,\ket {J=\ell}{S=0}m\nonumber\\[0.1in]
\ket {J=\ell}{j'=\ell+1/2}m & \sqrt{{J\over 2J+1}}\,\ket {J=\ell}{S=1}m  - \sqrt{{J+1\over 2J+1}}\,\ket {J=\ell}{S=0}m\nonumber\\[0.1in]
\ket {J=\ell-1}{j'=\ell-1/2}m & \ket {J=\ell-1}{S=1}m\\[0.15in]
\ket{J=\ell+1}{j=\ell+1/2}m&\ket{J=\ell+1}{j'=\ell+1/2}m\nonumber\\[0.1in]
\ket{J=\ell}{j=\ell-1/2}m &{1\over 2J+1}\ket {J=\ell}{j'=\ell-1/2}m+
{2\sqrt{J(J+1)}\over 2J+1}\ket {J=\ell}{j'=\ell+1/2}m  \nonumber\\[0.1in]
\ket{J=\ell}{j=\ell+1/2}m &{2\sqrt{J(J+1)}\over 2J+1}\ket {J=\ell}{j'=\ell-1/2}m
     -{1\over 2J+1}\ket {J=\ell}{j'=\ell+1/2}m \nonumber\\[0.1in]
\ket{J=\ell-1}{j=\ell-1/2}m &\ket{J=\ell-1}{j'=\ell-1/2}m\\[0.1in]
\end{array}\end{equation}
In these three bases it is easy to evaluate the matrix elements of 
${\bm \ell}\ncdot{\bm s}_1$, ${\bm \ell}\ncdot{\bm s}_2$
\begin{equation}
\begin{array}{@{\langle} l@{|\,}c@{\,|}r@{\rangle\ = \ } l}
Jjm    &2{\bm \ell}\ncdot{\bm s}_1& Jjm    &j(j+1)-\ell(\ell+1)-3/4\nonumber\\[0.1in]
J{j'}m &2{\bm \ell}\ncdot{\bm s}_2& J{j'}m &j'(j'+1)-\ell(\ell+1)-3/4
\end{array}
\end{equation}
and of $S_{12}$ (all of whose matrix elements vanish if the initial or
final state has $S=0$)\vspace{0.3in}
\begin{equation}
\begin{array}{@{\langle}l@{|S_{12}|}l@{\rangle\ = \ }l}
{J=\ell-1},{S=1},m&{J=\ell-1},{S=1},m & -2{J+2\over 2J+1}\nonumber\\[0.1in]
{J=\ell},{S=1},m &{J=\ell},{S=1},m  & +2\nonumber\\[0.1in]
{J=\ell+1},{S=1},m &{J=\ell+1},{S=1},m & -2{J-1\over 2J+1}\\[0.1in]
\end{array}
\end{equation}
from which we find
\begin{equation}
\begin{array}{@{\langle} l@{|\,}c@{\,|}r@{\rangle\ = \ } l}
J=\ell+1,\,j=\ell+1/2,m    &2{\bm \ell}\ncdot{\bm s}_2&J=\ell+1,\,j=\ell+1/2,m       &J-1\nonumber\\[0.1in]
J=\ell,\,j=\ell+1/2,m    &2{\bm \ell}\ncdot{\bm s}_2&J=\ell,\,j=\ell+1/2,m       &-{J(2J+3)\over 2J+1}\nonumber\\[0.1in]
J=\ell,\,j=\ell-1/2,m    &2{\bm \ell}\ncdot{\bm s}_2&J=\ell,\,j=\ell-1/2,m       &{(2J-1)(J+1)\over 2J+1}\nonumber\\[0.1in]
J=\ell,\,j=\ell-1/2,m    &2{\bm \ell}\ncdot{\bm s}_2&J=\ell,\,j=\ell+1/2,m       &-{2\sqrt{J(J+1)}\over 2J+1}\nonumber\\[0.1in]
J=\ell-1,\,j=\ell-1/2,m    &2{\bm \ell}\ncdot{\bm s}_2&J=\ell-1,\,j=\ell-1/2,m       &-J-2\non
\end{array}
\end{equation}
and
\begin{equation}
\begin{array}{@{\langle} l@{|\,S_{12}\,|}r@{\rangle\ = \ } l}
J=\ell+1,\,j=\ell+1/2,m    &J=\ell+1,\,j=\ell+1/2,m       &-2{J-1\over 2J+1}\nonumber\\[0.1in]
J=\ell,\,j=\ell+1/2,m    &J=\ell,\,j=\ell+1/2,m       &{2J\over 2J+1}\nonumber\\[0.1in]
J=\ell,\,j=\ell-1/2,m    &J=\ell,\,j=\ell-1/2,m       &{2(J+1)\over 2J+1}\nonumber\\[0.1in]
J=\ell,\,j=\ell-1/2,m    &J=\ell,\,j=\ell+1/2,m       &{2\sqrt{J(J+1)}\over 2J+1}\nonumber\\[0.1in]
J=\ell-1,\,j=\ell-1/2,m    &J=\ell-1,\,j=\ell-1/2,m       &-2{J+2\over 2J+1}\non
\end{array}
\end{equation}

\section*{Acknowledgment}
This work was supported in part by the Director, Office of Science, Office of High Energy and Nuclear Physics, of the U.S. Department of Energy under Contract
DE-AC0376SF00098.
\vspace{0.2in}
\end{widetext}

\end{document}